# Data-Driven Dispatchable Regions with Potentially Active Boundaries for Renewable Power Generation: Concept and Construction

Yanqi Liu, Zhigang Li, *Senior Member, IEEE*, Wei Wei, *Senior Member*, *IEEE*, J. H. Zheng, and Hongcai Zhang, *Member*, *IEEE*

*Abstract*—The dispatchable region of volatile renewable power generation (RPG) quantifies how much uncertainty the power system can handle at a given operating point. State-of-the-art dispatchable region (DR) research has studied how system operational constraints influence the DR but has seldom considered the effect of the uncertainty features of RPG outputs. The traditional DR is generally described by a large number of boundaries, and it is computationally intensive to construct. To bridge these gaps, a novel type of DR is defined, which is enclosed by potentially active boundaries (PABs) that consider the operational constraints and uncertainty features of RPG outputs. The proposed DR is easier to construct because the PABs are only a small part of the traditional DR boundaries. The procedure for constructing the proposed DR is described in terms of the progressive search for PABs, which is formulated as a mixed-integer linear program by incorporating the discrete observed data points of RPG outputs as an approximate distribution. A parallel solution paradigm is also developed to expedite the construction procedure when using a large observed dataset. Simulation tests on the IEEE 30-bus and 118-bus systems verify the effectiveness and scalability of the proposed DR and the efficiency of the proposed algorithm.

## NOMENCLATURE

The symbols used in this paper are listed below. Others are defined following their first appearance.

**Acronyms**

| | |
|---|---|
| Ad-CG | Adaptive constraint generation. |
| DHS | District heating system. |
| DNE | Do-not-exceed limit. |
| DR | Dispatchable region. |
| MILP | Mixed-integer linear program. |
| NAB | Never-active boundaries. |
| ODP | Observed data point. |
| PAB | Potentially active boundaries. |
| RPG | Renewable power generation. |

**Indices**

| | |
|---|---|
| $i$ | Index for conventional units. |
| $j$ | Index for renewable energy units. |
| $l$ | Index for transmission lines. |
| $q$ | Index for loads. |
| $I$ | Index set for potentially active boundaries. |
| $k$ | Index for observed data points of renewable power generation output. |
| $m$ | Index for subdatasets in the parallel method. |

**Parameters**

| | |
|---|---|
| $A$ | Coefficient matrix corresponding to the operational constraints. |
| $b_0$ | Coefficient vector corresponding to the operational constraints. |
| $b$ | Coefficient vector corresponding to the operational constraints. |
| $B$ | Coefficient matrix corresponding to the operational constraints. |
| $C$ | Coefficient matrix corresponding to the operational constraints. |
| $F_l$ | Active power capacity of transmission line $l$. |
| $M$ | Positive number in the Big M method. |
| $p_i$ | Basepoint output of conventional unit $i$. |
| $p^c_i$ | Corrective output of conventional unit $i$. |
| $p_q$ | Electric power load $q$. |
| $r_i^+$ | Ramp-up limit of conventional unit $i$. |
| $r_i^-$ | Ramp-down limit of conventional unit $i$. |
| $w^e_j$ | Forecasted output of renewable energy unit $j$. |
| $x$ | Base point of the predispatch strategy. |
| $\Delta w_k$ | The $k$th observed renewable power generation output. |
| $\Delta t$ | Time duration of the current dispatch interval. |
| $\pi_{il}$ | Power shift factor from conventional unit $i$ to line $l$. |
| $\pi'_{jl}$ | Power shift factor from renewable energy unit $j$ to line $l$. |
| $\pi''_{ql}$ | Power shift factor from load $q$ to line $l$. |

Y. Liu, Z. Li, and J. Zheng are with the School of Electric Power Engineering, South China University of Technology, Guangzhou 510641, China. W. Wei is with the Department of Electrical Engineering, Tsinghua University, Beijing 100084, China. H. Zhang is with the State Key Laboratory of Internet of Things for Smart City and Department of Electrical and Computer Engineering, University of Macau, Macao, 999078 China. (Corresponding author: Zhigang Li, lizg16@scut.edu.cn).

This work was supported by the National Natural Science Foundation of China (Grant No. 52177086), Guangdong Basic and Applied Basic Research Foundation (Grant No. 2019A1515011408), the Scientific and Technology Program of Guangzhou (Grant No.201904010215), and the Talent Recruitment Project of Guangdong (Grant No. 2017GC010467).

**Decision Variables**

$p_i^+$      Upregulation power of generator $i$.
$p_i^-$      Downregulation power of generator $i$.
$\boldsymbol{u}$      Dual variable associated with the primary constraints of system operation.
$z_k$      Binary auxiliary variable indicating whether the $k$th data point is separated from the dispatchable region.
$\Delta w_j$      Uncertain output of renewable energy unit $j$.

**Vector and Matrix Notation**

The vectors and matrices used in this paper are defined as follows: Vector $\boldsymbol{p}=\{p_i\}\ \forall i$, $\boldsymbol{p}^+=\{p_i^+\}\ \forall i$, $\boldsymbol{p}^-=\{p_i^-\}\ \forall i$, $\boldsymbol{w}^e=\{w_j^e\}\ \forall j$, and $\Delta\boldsymbol{w}=\{\Delta w_j\}\ \forall j$. Vector $\boldsymbol{x}$ denotes the predispatch strategy $\{\boldsymbol{p},\boldsymbol{r}\}$. Vector $\boldsymbol{y}$ represents the redispatch action $\{\boldsymbol{p}^+,\boldsymbol{p}^-\}$ after the uncertainty of the renewable energy is determined. Matrix $\boldsymbol{H}$ and vector $\boldsymbol{h}$ denote the polyhedral formulation of the dispatchable region.

I. INTRODUCTION

To reduce fossil fuel dependence, greenhouse gas emissions, and environmental pollution, renewable energy has received increased attention in the development of power systems. Because of its low generation cost and environmental impact, wind power generation has been prioritized in China, with national totals accounting for 35% of the world wind capacity [1]. However, the uncertainty and variability in wind power affect the safe and reliable operation of power systems.

To effectively utilize wind power, system operators need to allocate sufficient reserve capacity to address the uncertainty in power generation. Reference [2] modeled the power dispatch procedure as a two-stage process, including the predispatch and redispatch stages. Robust optimization was employed to consider wind generation uncertainty, and the feasibility of the optimal strategy was validated. The steady-state secure region, defined as the ranges of the nodal power injection limits, was utilized to describe the wind power generation limits in economic dispatch in [3]. Reference [4] used the concept of the regulating region of the district heating system (DHS) to describe the reserve flexibility that a single DHS can provide, which is a homologous formulation with the dispatchable region. These references employ the dispatchable region in different optimization problems and highlight the need to analyze the feasible region of generator outputs. It is valuable to study the dispatchable region of renewable power generation (RPG) in power system operation with RPG integration.

The security of the operation of a large-scale power system with volatile RPG integration has been investigated in recent studies. Reference [5] assessed the wind power provided by a power system under a unit commitment strategy and quantified the operational risk. Based on [5], a dynamic risk-based uncertainty set for decision making involving robust unit commitment was suggested in [6]. Furthermore, [7] investigated wind power utilization in integrated electric-gas systems employing the wind power admissible region in [5]. Additionally, following [5], the potential risks of integrating wind power and its maximum accommodation range with the help of an energy storage device were analyzed in [8].

Reference [9] defined the real-time dispatchable region (DR) as the range of wind power output that a power system can accommodate under given conditions and analyzed the feasibility of real-time dispatch with wind power integration. Based on the work of [9], the impact of wind power variations on operational constraints and the redispatch cost was discussed in [10], and an adaptive constraint generation (Ad-CG) algorithm was proposed to construct the dispatchable region. Reference [11] employed the Ad-CG algorithm to solve a joint robust economic dispatch problem for power systems incorporating wind power generation and carbon capture power plants. By utilizing the forecasted output and confidence intervals for renewable energy, [12] formulated a bilevel robust optimization model to assess the secure reserve level in power systems under uncertainties, and the worst-case scenarios were quantified.

Reference [13] proposed the concept of the do-not-exceed (DNE) limit and calculated the maximum range of RPG that the power system could accommodate during secure operation. Reference [14] proposed a unified framework for assessing system flexibility and the acceptable range of RPG outputs, and this approach was found to be applicable to system planning over long time scales. Reference [15] analyzed the flexibility of a power system by verifying the feasibility of robust multiperiod security-constrained optimal power flows with predefined unit commitments, and improvement measures were proposed to overcome insufficient flexibility issues. Reference [16] proposed an AC-power-flow-based DR model considering reactive power and voltage profile constraints. The inherently nonlinear AC power flow equations were approximately linearized using logarithmic transforms of the voltage magnitudes, and the quadratic constraints were approximated using a polytope approximation technique. Reference [17] estimated the secure operation region by treating nonlinear constraints as convex inner approximations. However, this approach tends to be overly conservative in large-scale power systems. To avoid the conservativeness of robust optimization, historical observed data were utilized to obtain probabilistic distributions for RPG outputs. References [18] and [19] formulated an optimization model with historical observed data to approximate the RPG distribution characteristics.

Most existing work on DR has solely focused on the effect of the system operational constraints on defining the DR, regardless of the uncertainty feature of the RPG outputs. Considering the uncertainty features of RPG, only some of the DR boundaries are potentially active, while the others are never active because no realistic RPG output scenarios will ever violate them. Hence, it is sufficient to identify only the potentially active boundaries (PABs) to determine the range of the dispatchable RPG outputs, which improves computational efficiency without losing practically useful information. Furthermore, traditional DRs are typically described by numerous boundaries. Enumerating all the boundaries, as in the traditional DR method, is computationally expensive when multiple RPG units are involved.

To address these issues, a data-driven DR for RPG with PABs is defined and constructed in this paper. First, the concept of the PAB is introduced by considering the impact of the uncertainty features of actual RPG outputs on the DR. Second, a DR incorporating PABs is formulated. The procedure for constructing the proposed DR is cast as a progressive search for PABs. The problem of updating PABs is formulated as a mixed-integer linear program (MILP) by adopting historical observed data to capture the empirical probabilistic information of RPG outputs. To reduce the computational burden associated with large-scale historical datasets, a parallel solution paradigm is developed to efficiently handle the boundary search problem. The contributions of this paper are threefold:

1) We define and formulate a DR with PABs. To the best of our knowledge, this is the first study to reveal the impact of RPG output uncertainties on the validity of the DR.

2) An iterative construction procedure for the proposed DR is developed. As the key problem, the search for PABs is formulated as a mixed-integer linear program by incorporating observed data points (ODPs) for RPG outputs. This problem can be readily solved using commercial solvers.

3) A parallel solution framework is proposed to efficiently expedite the DR construction procedure to accommodate large-scale historical observed datasets of RPG outputs.

The rest of this paper is organized as follows. In Section II, the mathematical model of the traditional DR is reviewed, and the limitations of the traditional DR are discussed. Section III presents the definition of a DR with PABs and develops the construction procedure for the proposed DR as well as its parallel solution strategy. In Section IV, numerical simulations are conducted to verify the effectiveness of the proposed DR and the efficiency of the proposed construction algorithm. Conclusions are drawn in Section V.

## II. THE DISPATCHABLE REGION AND ITS LIMITATIONS

### A. Formulation of the Dispatchable Region

A DR is a set of RPG outputs that will not cause infeasibility in power system operation during the redispatch process of conventional generation units. In fact, the DR describes the largest range of RPG that the power system can accommodate and characterizes the flexibility of the power system under the current operating point [10]. The larger the DR region is, the higher the level of uncertainty the power system can cope with.

In this context, the DR, denoted as $W^{\text{RTD}}$, can be formulated in the following abstract form:

$$W^{\text{RTD}} = \{\Delta \boldsymbol{w} \in \mathbb{R}^n \mid \exists \boldsymbol{y} \in \mathbb{R}_+^m, \\ \boldsymbol{By} + \boldsymbol{C}(\Delta \boldsymbol{w} + \boldsymbol{w}^e) \leq \boldsymbol{b}^0 - \boldsymbol{Ax}\} \quad (1)$$

where the actual output of renewable energy is the sum of the forecasted output $w^e$ and the forecasting error $\Delta w$. Because the forecasting error is unknown until the actual wind power output is observed, the redispatch actions $\{p^+, p^-\}$ need to be deployed after the actual output of the renewable energy is determined. $x$ represents the predispatch strategy, the generators' output and reserve are $\{p,r\}$, and $y$ represents the redispatch action $\{p^+,p^-\}$. $A$, $B$, $C$, and $b$ are the coefficient matrices or vectors corresponding to the operational constraints of power systems.

Specifically, the linear inequalities in (1) represent the following linear constraints:

$$p_i^c = p_i + p_i^+ - p_i^-, \quad \forall i, \quad (2)$$

$$\sum_i p_i^c + \sum_j (w_j^e + \Delta w_j) = \sum_q p_q, \quad \forall i, \quad (3)$$

$$-F_l \leq \sum_i \pi_{il} p_i^c + \sum_j \pi'_{jl}(w_j^e + \Delta w_j) \\ -\sum_q \pi''_{ql} p_q \leq F_l, \quad \forall l \quad (4)$$

$$0 \leq p_i^+ \leq r_i^+ \quad \forall i, \quad (5)$$

$$0 \leq p_i^- \leq r_i^- \quad \forall i, \quad (6)$$

where $\boldsymbol{p}$ is the basepoint output of conventional units and $\boldsymbol{r}$ represents their reserve capacity. $\boldsymbol{p}^c$ is the corrective output of conventional units in the redispatch process. $\boldsymbol{p}_q$ denotes the power of the loads. The subscript $i$ is the index of the conventional units, and $j$ is the index of the renewable energy units. The subscript $l$ is the index of the transmission lines. $\pi_{il}$ is the power shift factor from conventional unit $i$ to line $l$. $\pi'_{jl}$ is the power shift factor from renewable energy unit $j$ to line $l$. $\pi''_{ql}$ is the power shift factor from load $q$ to line $l$.

Equation (2) describes the relationship between the corrective output and the basepoint output. Equation (3) is the active power balance constraint at bus $i$. Equation (4) represents the power limits of the transmission lines. Equations (5) and (6) indicate that the corrective output cannot exceed the reserve capacity. Letting $\boldsymbol{b}=\boldsymbol{b}^0-\boldsymbol{Cw}^e$, the DR described by (1) can be expressed in the following form:

$$W^{\text{RTD}} = \{\Delta \boldsymbol{w} \in \mathbb{R}^n \mid \exists \boldsymbol{y} \in \mathbb{R}_+^m, \\ \boldsymbol{By} + \boldsymbol{C}\Delta \boldsymbol{w} \leq \boldsymbol{b} - \boldsymbol{Ax}\} \quad (7)$$

### B. Limitations of the Traditional Dispatchable Region

The DR described in (7) can be analytically expressed in the following polyhedral form [10]:

$$W^{\text{RTD}}(\boldsymbol{x}) = \{\Delta \boldsymbol{w} \mid \boldsymbol{u}_i^T (\boldsymbol{C}\Delta \boldsymbol{w} - \boldsymbol{b} + \boldsymbol{Ax}) \geq 0, \\ \forall i = 1, 2, ..., n\} \quad (8)$$

where $\boldsymbol{u}_1, \boldsymbol{u}_2, ..., \boldsymbol{u}_n$ are the vertices of the set $U=\{\boldsymbol{u}| \boldsymbol{B}^T\boldsymbol{u} = \boldsymbol{0}, -\boldsymbol{1}\leq\boldsymbol{u}\leq\boldsymbol{0}\}$. The DR is a polyhedron enclosed by a series of boundaries. Each boundary is a hyperplane that excludes the infeasible RPG outputs from the DR. The traditional DR enumerates all the boundaries of the polyhedron described in (8) and exposes them to the system operators.

The RPG outputs have specific uncertainty features determined by geographical and climate factors relating to the RPG plant location. A boundary is deemed to be *potentially active* only if its corresponding constraints may be violated by likely-to-happen RPG outputs. In reality, only a few of the boundaries of the traditional DR are potentially active, while the others are *never-active boundaries* (NABs). Consequently, it is sufficient to present only the PABs instead of showing all the DR boundaries.

A simple example is presented below to illustrate the limitations of the traditional DR. Fig. 1 shows the DR of two wind farms in a 5-bus power system. The horizontal and vertical coordinates denote the deviated outputs from the basepoints of the two wind farms. Each line in this figure represents a boundary related to the operational constraints of the power systems. The green region enclosed by all boundaries denotes the traditional DR.

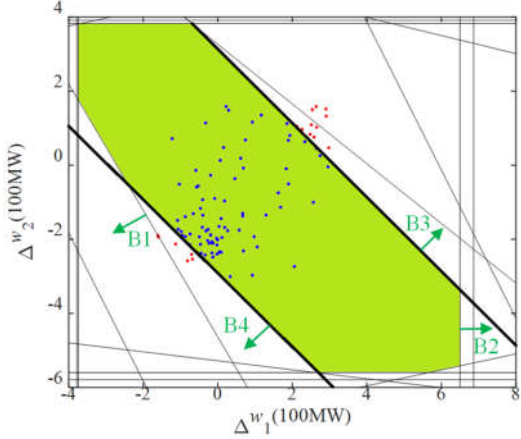

Fig. 1 The traditional DR and its boundaries

In this paper, the uncertainty features of the RPG output are not explicitly expressed using the probability distribution functions in closed form. Instead, they are implicitly described using the historical data points of RPG outputs in a data-driven fashion, which implicitly captures realistic patterns in the empirical distribution [20]. Each historical data point is referred to as an ODP of the RPG outputs, which can be seen as plausible RPG output scenarios according to the empirical distribution. The scattered points in Fig. 1 represent the ODPs for wind power output deviations, among which red and blue denote instances outside and inside the DR, respectively. The arrows on boundaries B1, B2, B3, and B4 indicate the directions of the corresponding boundaries that exclude the RPG output scenarios.

As shown in the figure, boundaries B1 and B2 do not exclude any of the ODPs of the RPG outputs, and boundaries B3 and B4 exclude all the infeasible ODPs, as the red points indicate. Boundaries B3 and B4 are the most informative to the system operators because they are PABs, i.e., they effectively exclude the infeasible ODPs of the RPG outputs. In contrast, boundaries B1 and B2 are NABs and are less informative to the operators because they merely consider the physical ability of the system to accommodate RPG outputs but neglect the actual probabilistic information of these outputs. The PABs provide instructive guidance to the system operators about the ability of the system to accommodate RPG with specific probabilistic information. In contrast, most of the boundaries of the traditional DR are NABs, and enumerating them requires a high computational burden. Therefore, traditional DR is limited in practical use.

### III. DISPATCHABLE REGIONS WITH POTENTIALLY ACTIVE BOUNDARIES

*A. Definition of Dispatchable Regions with Potentially Active Boundaries*

To address the aforementioned issues associated with the traditional DR, this paper proposes a DR for RPG with PABs. Mathematically, a DR boundary is potentially active if the set of RPG output scenarios violating this hyperplane is nonempty. The proposed DR is defined as follows:

$$\tilde{W}^{\text{RTD}}(\boldsymbol{x}) = \{\Delta\boldsymbol{w} \mid \boldsymbol{u}_i^T (\boldsymbol{C}\Delta\boldsymbol{w} - \boldsymbol{b} + \boldsymbol{A}\boldsymbol{x}) \geq 0, \forall i \in I\}, \quad (9)$$

where $I$ represents the index set of PABs:

$$I = \left\{ i \in \{1,2,...,n\} \middle| \exists \Delta\boldsymbol{w} \in \Omega, \boldsymbol{u}_i^T (\boldsymbol{C}\Delta\boldsymbol{w} - \boldsymbol{b} + \boldsymbol{A}\boldsymbol{x}) < 0 \right\}. \quad (10)$$

In (10), $\Omega$ denotes the set of ODPs of the RPG outputs. Equation (10) describes the index set of the PABs that exclude the infeasible ODPs of RPG outputs. Because $I \subseteq \{1,2,...,n\}$, we have $\tilde{W}^{\text{RTD}} \supseteq W^{\text{RTD}}$.

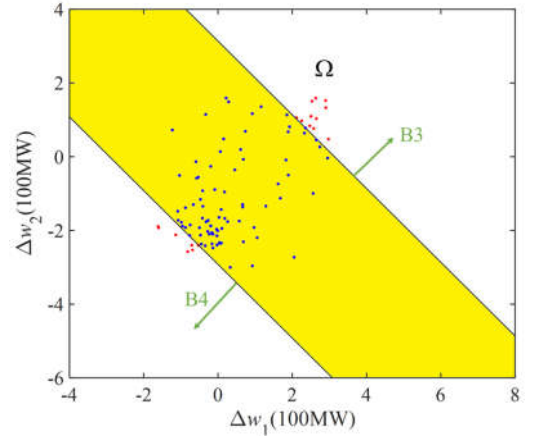

Fig. 2 The DR based on PABs

The set of scattered points in Fig. 1 is indeed an approximation of $\Omega$. For the example in Fig. 1, we have $I = \{3,4\}$, and the corresponding DR with PABs $\tilde{W}^{\text{RTD}}$ is shown as the yellow region in Fig. 2. The boundaries of $\tilde{W}^{\text{RTD}}$ include only B3 and B4, rather than all those in Fig. 1. It is obvious that $\tilde{W}^{\text{RTD}}$ in Fig. 2 contains $W^{\text{RTD}}$ in Fig. 1. Although $\tilde{W}^{\text{RTD}}$ has fewer boundaries, it separates out the same set of infeasible ODPs for the RPG outputs as does the traditional DR. Compared to the traditional DR, $\tilde{W}^{\text{RTD}}$ provides critical boundary information to system operators with fewer redundant boundaries, and the search process requires less computational effort, as will be shown in the case study.

*Discussion*: The PAB is essentially a type of DR boundary but is not the same as a DR boundary. The general DR boundaries are defined solely by the system's own abilities (e.g., physical and other operating constraints), while the PABs are restricted not only by the system's abilities but also by the uncertainty behavior of RPG outputs. In other words, the types of installed RPG units (e.g., photovoltaics, wind turbines, and biomass) do not have any impact on the general DR boundaries; however, they matter for the PABs because different patterns of RPG uncertainties will be involved. The PAB set is also a particular subset of the umbrella set [21] because the PAB is constrained not only by the system's ability but also by the uncertainty behavior of the RPG outputs. In summary, PABs can provide the most useful DR information with moderate computational effort.

*B. Progressive Construction of the DR with PABs*

This paper adopts a progressive construction algorithm to calculate the DR based on the PAB concept. The basic idea of this algorithm is similar to that of column-and-constraint generation [22], the convergence of which has been demonstrated. The key step in the proposed algorithm is to search for the PABs that exclude the most infeasible ODPs in each iteration. In this way, the PABs are generated progressively to construct the DR $\tilde{W}^{\text{RTD}}$ until all the infeasible ODPs are excluded from the DR. In detail, the proposed procedure includes the following steps:

**Step 1**: Initialization. Suppose the base point of dispatch is $x$. Let $\Omega = \{\Delta w_k: k = 1, 2, \ldots, n\}$ represent the historical ODPs of RPG outputs. Initialize a set $W^{\text{B}} = \{\Delta w \mid H\Delta w \geq h\}$ that is large enough to contain the final $\tilde{W}^{\text{RTD}}$. Set the value of the constant parameter $M$.

**Step 2**: Search for the next PAB. The MILP problem is a boundary search problem that aims to generate the PAB that excludes the most infeasible ODPs in the current iteration.

$$v = \max_{u,z} \sum_k z_k \quad (11)$$

Subject to
$$B^T u = 0, \quad (12)$$

$$-1 \leq u \leq 0, \quad (13)$$

$$z_k \in \{0,1\}, \; \forall \{k: \Delta w_k \in \Omega \cap W^{\text{B}}\}, \quad (14)$$

$$z_k = 0, \; \forall \{k: \Delta w_k \notin \Omega \cap W^{\text{B}}\}, \quad (15)$$

$$(C\Delta w_k - b + Ax)\cdot u + z_k \cdot M \leq M, \\ \forall \{k: \Delta w_k \in \Omega \cap W^{\text{B}}\}, \quad (16)$$

$$(C\Delta w_k - b + Ax)\cdot u + z_k \cdot M \geq 0, \\ \forall \{k: \Delta w_k \in \Omega \cap W^{\text{B}}\}, \quad (17)$$

where the decision variables are $u$ and $z$. The vector of variables $u$ represents the dual variable associated with the primary constraints in (7) and the normal vectors of the boundaries, which are constrained by (12)-(13). The vector of variables $z$ contains binary auxiliary variables. Equations (14) and (15) show that $z_k$ is binary if the $k$th data point is inside the current $W^{\text{B}}$ and zero otherwise. For $z_k$ inside the current $W^{\text{B}}$, $z_k = 1$ implies that the $k$th data point is separated from the DR by the hyperplane $u_i^T(C\Delta w - b + Ax) = 0$, and $z_k = 0$ implies the opposite, as stipulated in (16)-(17). Hence, the auxiliary variable $z_k$ indicates the positional relation between the $k$th ODP for the RPG outputs and the hyperplane $u_i^T(C\Delta w - b + Ax) = 0$ to be determined.

The objective of the boundary search problem is to maximize the number of infeasible ODPs that are separated from the DR by the hyperplane to be determined. The solution to this problem provides the normal vector $u^*$ of the hyperplane that excludes the largest number (namely, $v^*$) of infeasible ODPs for the RPG outputs in the current iteration. As a MILP problem, the boundary search problem can be solved by commercial solvers.

**Step 3**: Update $W^{\text{B}}$. If $v^* > 0$, which indicates that there are infeasible ODPs inside $W^{\text{B}}$, a new PAB is included to update the set $W^{\text{B}} := W^{\text{B}} \cap \{\Delta w \mid u^{*T}(C\Delta w - b + Ax) \geq 0\}$, and the algorithm returns to Step 2. If $v^* = 0$, implying that all infeasible ODPs have been excluded from $W^{\text{B}}$, let $\tilde{W}^{\text{RTD}} = W^{\text{B}}$ and terminate the algorithm.

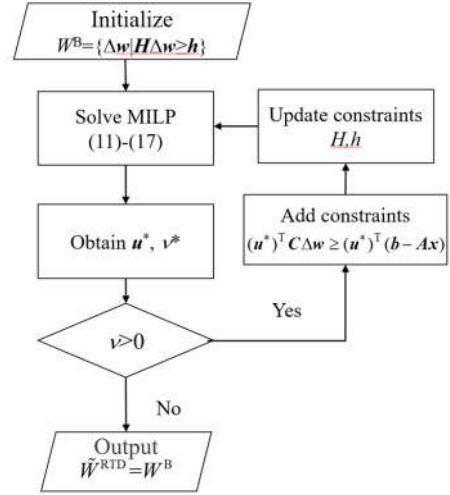

Fig. 3 Flowchart of the progressive construction procedure

The proposed construction procedure is different from the traditional procedure in [10]. The latter aims to search all boundaries of the DR subject to only the operational constraints in (2)-(6). In contrast, the proposed algorithm identifies PABs of the DR that not only consider operational constraints but also exclude the infeasible ODPs of RPG outputs. Specifically, the boundary search problem aims to search for PABs that exclude the largest number of infeasible ODPs in each iteration. Compared to the construction procedure in [10], the proposed approach returns many fewer boundaries while considering the actual distributional characteristics of the RPG outputs, and hence, the computation time can potentially be reduced.

*C. Parallel Computation Framework of the Construction Procedure*

The actual distributional characteristics of the RPG outputs are considered by using historical data in the proposed procedure. According to the law of large numbers, a large dataset is needed to accurately approximate the empirical probabilistic information of the RPG outputs using discrete ODPs. Thus, a high computational burden is encountered in the proposed construction procedure because the complexity of solving the MILP problem in (11)-(17) grows exponentially with the number of binary variables.

To reduce the computation time, we propose a parallel computational framework for constructing the DR with PABs. The basic idea is to decompose the historical dataset $\Omega$ into multiple subsets; then, the PABs are searched in a parallel manner with each of these subsets before the DR is finally obtained by collecting all the resultant PABs.

The flowchart in Fig. 4 illustrates the parallel method. The calculation in each subdataset follows the construction procedure described in Fig. 3. The boundaries from each subdataset are shared with each other in the parallel calculation. The boundaries generated in the last iteration with a subdataset can help remove infeasible data points within the other subdataset in the next iteration of calculation. Compared with the nonparallel version in Fig. 3, the parallel method adds and utilizes the shared constraints at every iteration in the subdataset calculation. Then, the shared constraints can be filtered to exclude infeasible ODPs from the subdataset and to accelerate the computing process of boundary search in the next iteration.

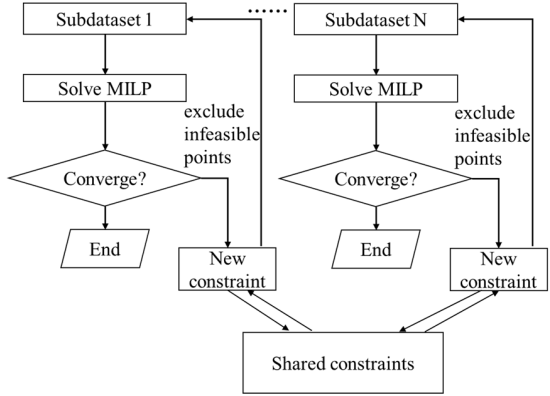

Fig. 4 Flowchart of the parallel method

The detailed steps in this framework are as follows:

**Step 1**: Decomposition of the dataset. Divide the historical dataset of RPG outputs $\Omega$ into $m$ subdatasets, $\Omega_1, \Omega_2, \ldots, \Omega_m$, such that $\Omega = \Omega_1 \cup \Omega_2 \cup \cdots \cup \Omega_m$.

**Step 2**: Construction of DRs with subdatasets. With each subdataset $\Omega_k$ ($k = 1,2, 3, \ldots, m$), construct the corresponding DR with PABs, $W^B(k)$ ($k = 1, 2, \ldots, m$), according to the progressive construction procedure in Section III-B.

**Step 3**: Finalization. Collect $W^B(k)$ ($k = 1, 2, \ldots, m$) to obtain the final result for the DR with PABs by letting $\tilde{W}^{\text{RTD}} = W^B(1) \cap W^B(2) \cap \ldots \cap W^B(m)$.

In Step 2, constructing DRs with different subdatasets can be performed in a fully parallel manner. The total computational complexity of solving all boundary search problems with different subdatasets is $O(m2^{n/m})$, which is significantly reduced from that for the original dataset, i.e., $O(2^n)$. Therefore, the proposed parallel scheme can improve the efficiency of constructing DRs with a large set of ODPs.

IV. CASE STUDY

The proposed DR is compared with the traditional DR in terms of accuracy and computational efficiency. The basepoint of dispatching is calculated by MATPOWER [23], and the value of $M$ is set to 20000.

Case 1: A modified IEEE 30-bus power system [16] with two RPG units at buses #1 and #22. The forecasted outputs of RPG are 10 MW and 20 MW. The penetration rate of RPG is 15.86% in this case. The total load demand is scaled up to 189.20 MW. The other data are from the original IEEE 5-bus system.

Case 2: A modified IEEE 118-bus power system [16] with two RPG units at buses #70 and #49. This system consists of 54 conventional generators and 186 transmission lines. The total load demand is 5500 MW, and the forecasted RPG outputs are both 350 MW. The penetration rate of RPG is 12.73% in this case. The other data are collected from the original IEEE 118-bus system.

Case 3: Based on Case 2, an additional wind farm connects to bus #100 with a forecasted output of 200 MW. The penetration rate of RPG is 16.36% in this case.

Case 4: Based on Case 3, one more wind farm is added at bus #82, and the forecasted outputs of all wind farms are 150 MW. The penetration rate of RPG is 10.91% in this case.

Case 5: A modified IEEE 118-bus power system. Ten renewable resources are located at buses #4, #10, #12, #25, #31, #49, #59, #70, #82, and #100. The forecast RPG outputs are all set to 100 MW. The penetration rate of RPG is 18.18% in this case. The other data are collected from the original IEEE 118-bus system.

*A. Procedure for Constructing the DR*

The procedures for constructing the DR with the traditional and the proposed methods are compared. Specifically, the simple example shown in Fig. 5 is used to illustrate how DRs are constructed progressively in the proposed nonparallel procedure.

1) Input 500 historical data $W_0=\{\Delta w_k, k=1,2,\ldots,500\}$ of RPG at buses #1 and #22. The initial set $W^B$ is set as $W^B=\{\Delta w \mid [1\ 0;\ 0\ 1;\ -1\ 0;\ 0\ -1]\Delta w \geq -1000*[1;\ 1;\ 1;\ 1]\}$.

2) Put the parameters of the test system into constraints (2)-(6) and obtain the coefficient matrices $A$, $B$, $C$, and $b$, as shown in (1).

3) Solve the MILP problem presented in (11)-(17) with CPLEX to obtain the optimal value $v^*$ and the optimal solution $u^*$. In the first iteration, $v^*=28$ represents the 28 points outside the DR, and $u^*$ is used to form boundary B5, i.e., [-0.0504 -0.5368] $\Delta w \geq$ -0.1806 in Fig.5. Then, the rows [-0.0504 -0.5368] and [-0.1806] are added to $H$ and $h$, respectively.

4) The objective $v^*=28>0$ means that the PABs violated by the given historical data are not all found. Hence, return to (2) and start the next iterations, which generate boundaries B6, B7 and B8 sequentially.

Fig. 5 displays the DR with four PABs in Case 1. The historical ODPs are marked as blue or red points in Fig. 5 and Fig. 6. The blue points represent the ODPs inside the DR, and the red points denote those outside the DR. The four PABs, namely, B5 through B8, are generated sequentially. The first generated boundary B5 ($-0.0504\Delta w_1-0.5368\Delta w_2 \geq -0.1806$) excludes 28 infeasible ODPs in red from $W^B$. Boundary B6 ($\Delta w_1+0.0910\Delta w_2 \geq -0.8609$) is then generated to exclude two infeasible ODPs. The third generated boundary B7 ($-0.0910\Delta w_1-0.0910\Delta w_2 \geq -0.0741$) excludes the last two infeasible ODPs. The final generated boundary B8 ($-0.0264\Delta w_1-0.4601\Delta w_2 \geq -0.1396$) excludes two infeasible ODPs. It is provable that the number of newly excluded infeasible data points decreases to zero as iteration continues, implying the convergence of the proposed algorithm. This relation indicates that the PABs are generated based on the sequence of their ability to exclude the infeasible ODPs of the RPG outputs.

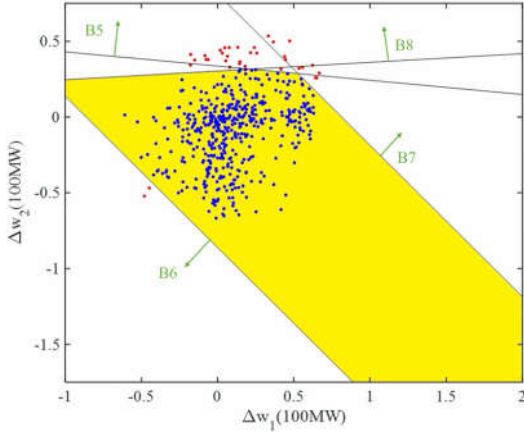

Fig. 5. The proposed DR with 500 historical data points for the RPG outputs in Case 1

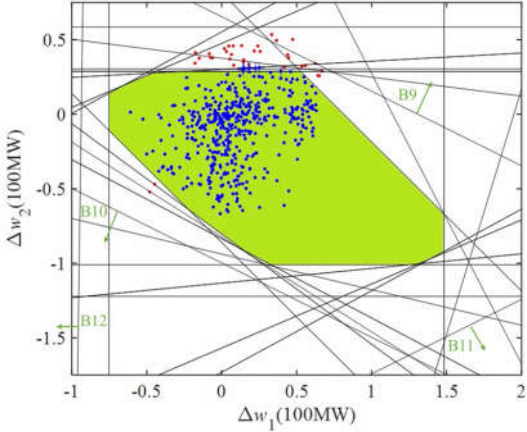

Fig. 6. The traditional DR with 500 historical data points for the RPG outputs in Case 1

Fig. 6 shows the traditional DR in green with 28 boundaries in total. B9, B10, B11 and B12 correspond to the first four boundaries that are generated successively. The first boundary excludes three infeasible ODPs. However, the next three boundaries are not potentially active because no ODPs are excluded. This is because the traditional DR construction method does not consider the probabilistic information of the RPG outputs, and all boundaries are treated equally.

The proposed DR and traditional DR results in Case 2 are shown in Fig. 7 and Fig.8, respectively. Fig.7 shows the DR with only one PAB, B13 ($-0.0228\Delta w_1 - 0.3072\Delta w_2 \geq -1.4402$), which excludes 10 infeasible ODPs in red from $W^B$. In contrast, the first two boundaries of the traditional DR, as shown by B14 and B15 in Fig. 8, fail to identify the infeasible ODPs and are thus NABs.

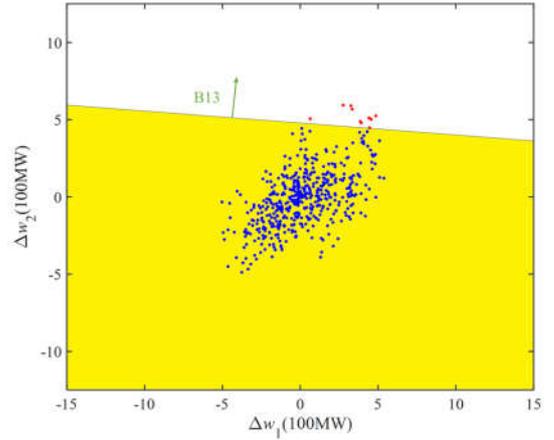

Fig. 7. The proposed DR with 500 historical data points for the RPG outputs in Case 2

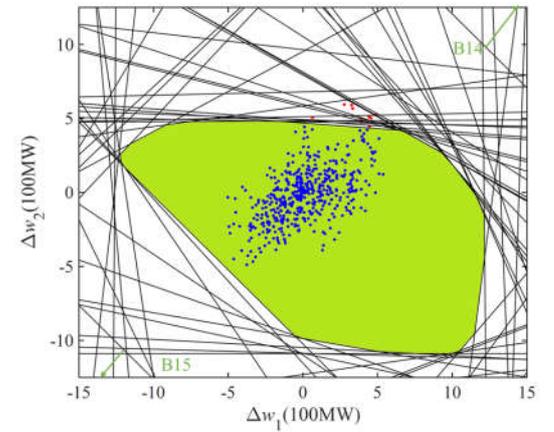

Fig. 8. The traditional DR with 500 historical data points for the RPG outputs in Case 2

As shown by the results, the proposed algorithm guarantees that the PAB generated at each iteration is always the most informative obtained thus far; i.e., it excludes the most infeasible ODPs at the current iteration. In contrast, the most informative boundaries do not have this priority during the construction of the traditional DR. In fact, the sequence of generating boundaries is particularly important when the computation time for constructing the DR is limited, as will be shown in Section IV-D.

B. *Computational Performance in Constructing DRs*

The computational performance in constructing different types of DRs is tested in this subsection.

For each case in TABLE I, Traditional refers to the procedure of constructing the traditional DR with the Ad-CG algorithm in [9]. Proposed-nPrl refers to the proposed nonparallel method, as given in Section III-B. Proposed-Prl refers to the proposed parallel method, as given in Section III-C. TABLE I shows that Proposed-nPrl converges in fewer iterations than Traditional. In Cases 1 and 2, with a small number of RPG units, both Proposed-Prl and Proposed-nPrl take longer than Traditional. When the number of RPG units increases in Cases 3 through 5, these two methods consume considerably less time than Traditional. This is because the number of DR boundaries increases exponentially with respect to the number of RPG units, while the number of PABs does not. In addition, a small number of PABs are generated by Proposed-nPrl because the operational constraints are violated for only a few ODPs in the historical dataset.

TABLE I
COMPUTATIONAL PERFORMANCE OF DIFFERENT METHODS

| Case | Method | Computational time (s) | # Boundaries found |
|---|---|---|---|
| 1 | Traditional | 0.96 | 28 |
|  | Proposed-nPrl | 8.30 | 4 |
|  | Proposed-Prl | 1.01 | 7 |
| 2 | Traditional | 10.21 | 85 |
|  | Proposed-nPrl | 28.87 | 3 |
|  | Proposed-Prl | 13.20 | 6 |
| 3 | Traditional | 135.23 | 164 |
|  | Proposed-nPrl | 76.61 | 5 |
|  | Proposed-Prl | 13.44 | 8 |
| 4 | Traditional | 385.8199 | 201 |
|  | Proposed-nPrl | 17.05 | 2 |
|  | Proposed-Prl | 12.61 | 7 |
| 5 | Traditional | 7248.70* | 236 |
|  | Proposed-nPrl | 65.91 | 3 |
|  | Proposed-Prl | 15.70 | 6 |

Notation: * terminated without convergence

We also compare the computational performance of Proposed-Prl and Proposed-nPrl, as shown in TABLE I. The total computation time of Proposed-Prl is significantly less than that of the nonparallel approach because the computational complexity of Proposed-Prl is remarkably reduced compared to that of the original procedure. Additionally, solving these subproblems simultaneously can effectively reduce the computational time. Moreover, the total number of PABs for both methods is identical because each subproblem in Proposed-Prl handles only a subset of the original dataset. However, after merging all the $W^B(k)$ ($k = 1, 2, …, m$) values, Proposed-Prl yields the same DR as Proposed-nPrl.

TABLE II
COMPUTATIONAL PERFORMANCE OF THE PROPOSED-NPRL METHOD WITH DIFFERENT NUMBERS OF HISTORICAL DATA

| $K_0$ | Case | Computational time (s) | Number of iterations |
|---|---|---|---|
| 100 | 1 | 0.3519 | 3 |
|  | 2 | 2.5594 | 2 |
|  | 3 | 3.3674 | 3 |
|  | 4 | 2.5773 | 2 |
|  | 5 | 3.08 | 2 |
| 250 | 1 | 1.6292 | 3 |
|  | 2 | 11.8062 | 3 |
|  | 3 | 13.1893 | 4 |
|  | 4 | 6.9539 | 2 |
|  | 5 | 11.4667 | 3 |
| 500 | 1 | 8.3078 | 4 |
|  | 2 | 29.378 | 3 |
|  | 3 | 77.4564 | 5 |
|  | 4 | 17.7509 | 2 |
|  | 5 | 66.6867 | 3 |
| 1000 | 1 | 21.9656 | 4 |
|  | 2 | 247.71 | 4 |
|  | 3 | 767.2768 | 8 |
|  | 4 | 132.5405 | 2 |
|  | 5 | 667.5517 | 5 |

## C. Impact of the ODP Number on the Computational Performance

The convergence performance and computational efficiency of constructing DRs with varying numbers of ODPs is discussed in this section. The results of Case 1 through Case 5 are displayed in TABLE II.

Fig. 9 shows that the computation time increases in a superlinear manner with the number of ODPs. As shown by the results, the computation time increases with the number of RPG units and the scale of the system. Notably, the number of integer variables increases exponentially with the number of RPG units. However, the number of iterations does not change significantly with variations in $K_0$ because solving the MILP problem accounts for the majority of the computational burden, and the number of integer variables is affected by the number of data points but not the scale of the system.

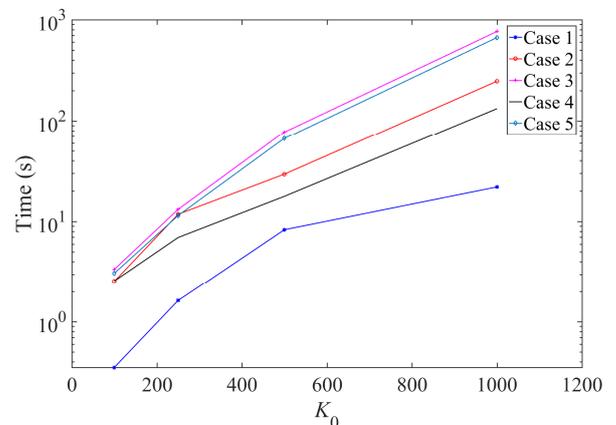

Fig. 9. Computational time of the proposed parallel method versus the number of historical data points

## D. Application of the Proposed DR in Real-Time Dispatch

One possible application of the DR is real-time power dispatch with RPG integration. The outputs of RPG are expected to be maximized as long as they do not incur security or stability issues, and the DR provides an effective tool to align real-time RPG units quickly. In real-time dispatch, the operating base points of all units are adjusted every five minutes [25]. Accordingly, the available computation time for obtaining DRs is restricted by this execution period; i.e., the procedure of updating the DR has to be manually terminated before it consumes more time than required. It is therefore desirable that the most informative boundaries of the DR can be found at the earliest stage, which is a distinguishing feature of the proposed method.

This subsection numerically shows the performance of the proposed DR over the traditional DR in real-time dispatch by using Case 5, the settings of which are the same as those given in Section IV-B. Five hundred ODPs from the NREL dataset [24] are utilized to test the effectiveness of these two regions. The time limit for DR calculation is set to three minutes.

The traditional and proposed DRs are generated. The comparison between these two DRs is listed in TABLE III. The calculation of the traditional DR is terminated in three minutes without final convergence, and 56 boundaries are found. The region encompassed by these 56 boundaries is referred to as Trad-DR, which includes all 500 ODPs. For each of the ODPs enclosed by this region, the feasibility check problem, as described in Appendix I, is solved to determine whether this point is feasible or not and to evaluate the constraint violation when it is infeasible. As TABLE III shows, 52 of the 500 ODPs enclosed by Trad-DR are infeasible, indicating an infeasibility rate of $52/500 \times 100\% = 10.40\%$, and the average constraint violation over these 500 ODPs is $9.72 \times 10^{-2}$. This means that if Trad-DR is employed for adjusting the RPG output in real-time dispatch, the chance of this leading to insecure operation is approximately 10.40%, which may not be acceptable in practical use.

TABLE III
COMPARISON BETWEEN THE TRADITIONAL DR AND THE PROPOSED DR IN REAL-TIME DISPATCH

| Region | Trad-DR | Proposed-DR |
|---|---|---|
| Computational time (s) | 180* | 65.91 |
| # Boundaries found | 56 | 3 |
| # Enclosed ODPs | 500 | 449 |
| # Infeasible ODPs | 52 | 1 |
| Infeasibility rate (%) | 10.40 | 0.67 |
| Average constraint violation | $9.72 \times 10^{-2}$ | $7.13 \times 10^{-6}$ |

Notation: * terminated without convergence

Without using the parallel method, the proposed DR is successfully constructed in 65.91 s, and the region encompassed by the three boundaries found is referred to as Proposed-DR. Although it consists of only three PABs, much fewer than the boundaries obtained by Trad-DR, Proposed-DR has a much lower infeasibility rate and significantly smaller average constraint violation than Trad-DR. This implies that Proposed-DR performs more reliably in determining the RPG output in real-time dispatch with less computational effort.

This advantage of Proposed-DR is due to its ability to generate the most informative boundary at each iteration, as analyzed in Section IV-A. When they are generated by the proposed method, the most informative boundaries are provided to the users before the less informative boundaries until the algorithm is manually terminated. In contrast, the traditional method does not follow this rule when generating boundaries. This explains why Trad-DR is not reliable in time-restricted applications, although a number of boundaries are found.

## V. CONCLUSIONS

This paper proposes a DR for RPG based on PABs. The PAB concept is defined for the first time to establish the relation between the boundaries of a DR and the empirical probabilistic information of RPG. A progressive construction algorithm is utilized to build the proposed DR with PABs, and the PABs that exclude the most infeasible ODPs of the RPG outputs are progressively generated through multiple iterations. To reduce the computational time associated with solving the boundary search problem, typically formulated as a large-scale MILP problem, we propose a parallel computational framework based on the idea of decomposing the historical dataset of RPG outputs. The numerical simulations illustrate the accuracy and computational efficiency of the proposed DR model and the corresponding construction approach. By considering the actual probabilistic information of RPG outputs, the number of boundaries required to build the DR decreases significantly. In addition, the most informative boundaries have priority in being generated in DR construction by the proposed method, making this method applicable for time-limited applications. The proposed parallel computational framework can efficiently construct the proposed DR, even with a large set of ODPs.

Several interesting research directions are now open based on the proposed DR. The sensitivity of the PABs to the historical data points is worth calculating to evaluate the robustness of the PAB-based DR to RPG uncertainties. The proposed PAB can also be employed to assess vulnerability or help in expansion planning for existing power networks, which will be an important part of our future work.

## APPENDIX I

Formulation of the Feasibility Check Problem

The following problem is defined to check whether there is a feasible solution $y$ that satisfies the constraints in (7) given an ODP $\Delta w^*$:

$$s(\Delta w^*) = \min_{\lambda, y} \mathbf{I}^T \lambda \tag{18}$$

$$\text{s.t.} \quad \mathbf{B} y - \lambda \leq \mathbf{b} - \mathbf{A} x - \mathbf{C} \Delta w_* \tag{19}$$

$$y \geq \mathbf{0}, \lambda \geq \mathbf{0} \tag{20}$$

The value of $s(\Delta w^*)$ can be regarded as a metric to indicate the minimum constraint violation given $\Delta w^*$. If $s(\Delta w^*) = 0$, there is a feasible solution to constraint set (7). $s(\Delta w^*) > 0$ implies that $\Delta w^*$ incurs infeasibility with respect to constraint set (7).

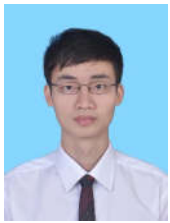

**Yanqi Liu** received his B.S. and M.S. degrees from the School of Electric Power engineering, South China University of Technology, Guangzhou, China. Currently, he is pursing the Ph.D. degree at University of Macau. His research interests include resilience assessment, renewable energy and integrated energy systems planning and operation.

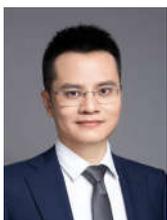

**Zhigang Li** (M'16-SM'21) received the Ph.D. degree in electrical engineering from Tsinghua University, Beijing, China, in 2016. He is currently an Associate Professor with the School of Electric Power Engineering, South China University of Technology, Guangzhou, China. He was a Visiting Scholar with the Illinois Institute of Technology, Chicago, IL, USA, and Argonne National Laboratory, Argonne, IL, USA. He serves as a Subject Editor for the CSEE Journal of Power and Energy Systems, and an Associate Editor for the Journal of Modern Power Systems and Clean Energy. His research interests include smart grid operation and control, integrated energy systems, optimization theory and its application in energy systems.

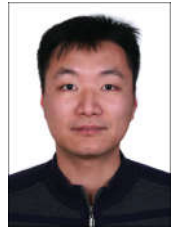

**Wei Wei** (SM'18) received the B.Sc. and Ph.D. degrees in electrical engineering from Tsinghua University, Beijing, China, in 2008 and 2013, respectively. He was a Postdoctoral Research Associate with Tsinghua University from 2013 to 2015. He was a Visiting Scholar with Cornell University, Ithaca, NY, USA, in 2014, and a Visiting Scholar with Harvard University, Cambridge, MA, USA, in 2015. He is currently an Associate Professor with Tsinghua University. His research interests include computational optimization and energy system economics.

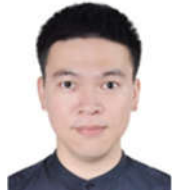

**J. H. Zheng** obtained his B.E. degree in Electrical Engineering from Huazhong University of Science and Technology, Wuhan, China, in 2012, and his Ph.D degree at the same area in South China University of Technology (SCUT), Guangzhou, China in 2017. He is currently a Lecturer in SCUT. His research interests include the area of optimization algorithms, decision making methods and their applications on integrated energy systems. He has authored or co-authored more than 40 peer-reviewed SCI journal papers.

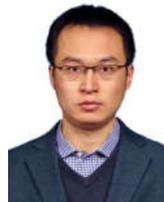

**Hongcai Zhang** (S'14-M'18) received the B.S. and Ph.D. degrees in electrical engineering from Tsinghua University, Beijing, China, in 2013 and 2018, respectively. He is currently an Assistant Professor with the State Key Laboratory of Internet of Things for Smart City and Department of Electrical and Computer Engineering, University of Macau, Macao, China. In 2018-2019, he was a postdoctoral scholar with the Energy, Controls, and Applications Lab at University of California, Berkeley, where he also worked as a visiting student researcher in 2016. His current research interests include Internet of Things for smart energy, optimal operation and optimization of power and transportation systems, and grid integration of distributed energy resources.